# On the equations of diffracted geodesics

D. H. Delphenich (†)
Spring Valley, OH 45370 USA

———

**Abstract**: An explicit form for the geodesic equations that would describe diffracted light rays is obtained and the Levi-Civita connection that enters into it is shown to be a sum of contributions from the metric of the ambient space, the index of refraction of the optical medium, and diffraction effects. The nature of the problem of obtaining explicit forms for the diffraction correction in various problems of interest to optics is discussed briefly.

**Keywords**: Geodesics, geometrical optics, wave optics, diffraction, geometrical theory of diffraction.

**1. Introduction.** – There are two basic approaches to the mathematical modelling of problems in optics that are referred to as wave optics and geometrical optics. In wave optics, the fundamental objects are wave surfaces that propagate through optical media that can change the shape and speed of propagation of the surfaces. In geometrical optics, the fundamental objects are light rays, which are presumed to be normal to the wave surfaces.

Actually, it was the latter that predated the former, since idea of modelling the behavior of light by lines goes back to at least the time of Euclid in ancient Greece, and was developed considerably by the Mesopotamian scientist Ibn al-Haytham (a.k.a., al Hazen) in the Eleventh Century A.D. Of course, for the Western World, it was most definitively established by Isaac Newton in 1730. Somewhat earlier, in 1662 Pierre de Fermat proposed his principle of least time as the basis for light rays, which formulated the problem of finding light rays in a medium as a problem in the calculus of variations. Thus, in their most modern manifestation, the equations of light rays become equations of geodesics for a definition of arc-length (viz., optical path length) that is based upon a conformal transformation of the metric on the ambient space in which the optical medium exists for which the conformal factor is the square of the index of refraction of the medium (at least for isotropic media.)

A few decades before Newton published his landmark treatise on ray optics, Christiaan Huygens published an even more far-reaching treatise on light in 1690 that proposed the wave-like nature of light. Due to the fact that Newton was far more established as a figure of science in that era, the work of Huygens went largely unnoticed for over a hundred years, at which point, Thomas Young, in 1809, and Augustin-Jean Fresnel, in 1827, resurrected it and developed it even further. The real crowning achievement of wave optics was in the late Nineteenth Century when James Clerk Maxwell and Heinrich Hertz established that light was basically one form of electromagnetic waves.

Meanwhile, largely due to the work of Sir William Rowan Hamilton in the early Nineteenth Century, a close analogy between the rays that are traced out by light in optical media and the

———

(†) E-mail: feedback@neo-classical-physics.info, Website: neo-classical-physics.info.

trajectories of massive points in the presence of external force fields was being developed. Indeed, Hamilton's equations of mechanics were actually first developed by him in the context of geometrical optics, so in a sense, he was applying an optical approach to the description of motion. That optical-mechanical analogy was later used as the basis for the development of quantum wave mechanics, which deals with matter waves as a representation of massive matter, by Louis de Broglie and Erwin Schrödinger.

One must really regard wave optics as something that is more physically definitive than geometrical optics, since light is most fundamentally wave-like in the eyes of electromagnetism. It is then governed by a system of partial differential equations, whereas geometric optics, which deals with curves in space deals with a system of ordinary differential equations. Although in principle as long as the wave surfaces of wave optics admit normal vector fields at every point, their time evolution should sweep out a corresponding congruence of light rays, in fact, the traditional treatment of geometrical optics introduces an approximation that amounts to reducing that congruence to a single light ray, or at least an "infinitely-thin" congruence of them. The information that is lost in that approximation amounts to the diffraction effects that are experienced by the wave surfaces as they propagate past optically-opaque obstacles, such as when they propagate through slits and apertures or around discs. However, one sees that in principle, light rays, as normal congruences to propagating wave surfaces, should not have to suffer such a loss of information. The reason for the introduction of the geometrical optics approximation seems to be mostly based in mathematical simplicity, rather than a fundamental fact of nature.

There have been attempts to remedy the oversight in geometrical optics by developing a "geometrical theory of diffraction" that would include the possibility of "diffracted geodesics." Notably, Joseph Keller published several articles on the topic in 1958 and 1962 [**1**]. However, the approach that he took was more heuristic than analytical, so in particular, although he suggested some possible characteristics of diffracted geodesics and a possible extension of Fermat's principle, he did not actually propose an extension of the geodesic equations that would include a contribution from diffraction near opaque boundaries.

The present study is an attempt to make some progress in the direction of extending the geodesic equations to include the effects of diffraction. One finds that the only difference between the equations that include diffraction effects and the classical equations of light rays in a linear homogeneous medium is an additive contribution to the conformal factor (i.e., the square of the index of refraction) that is a function of the amplitude of the light wave and will vanish in the limit of high wave number.

Although there is a crucial difference between the propagation of scalar waves, such as one finds in acoustics or elementary quantum mechanics, and the propagation of waves that are represented by vector fields (or differential 2-forms), such as electromagnetic waves, nonetheless, since the basic features of the diffraction effects on a congruence of light rays with a finite cross-section can be found in the simpler case of scalar waves, that will define the scope of the present development.

Ultimately, one finds that since obtaining a specific form for the contribution to the geodesic equations from diffraction involves knowing a specific form for the solution of the Helmholtz equation for a given optical configuration, and finding such a solutions is not as straightforward as it sounds, one must revert to more advanced techniques, such as asymptotic expansions, in order to make further progress. However, that would go beyond the scope of the present article, which seeks only to formulate the equations of diffracted geodesics for the propagation of scalar waves and examine their basic structure.

The next section addresses the separation of the time and space variables for a linear scalar wave equation when the speed of propagation of the waves varies with position and then the representation of the complex stationary wave function in polar coordinates, which defines its amplitude and phase functions. Section **3** then introduces the geometrical optics approximation in its usual form and shows how one can define a more general eikonal equation in which the index of refraction is perturbed by the amplitude of the wave. In section **4**, the geodesic equations for light rays in the medium that is governed by that perturbed index of refraction are obtained and decomposed into a part that is due to the metric on the ambient space in which the optical device exists, a part that is due to the unperturbed index of refraction of the optical medium, and a part that originates in the perturbation that is caused by diffraction. In section **5**, the nature of the problem of finding an explicit form for the amplitude of the wave function that could be used to obtain the perturbation of the geodesics by diffraction is discussed, and in the last section the results are summarized while some suggestions for promising directions of research to pursue are proposed.

The methodology of this article overlaps considerably with a previous one by the author [**2**] on the subject of geodesics in gradient-index media and the optical-mechanical analogy.

**2. Time-harmonic solutions of the linear scalar wave equation with variable propagation speed.** – The type of wave motion that we shall deal with here is perhaps the second-simplest case, in which one has a complex-valued scalar wave function $\psi(t, x^i)$ that describes a wave that propagates in a linear, but inhomogeneous, medium with a speed of propagation $v(x)$ that varies with position in space. Hence, in the absence of sources, $\psi$ will satisfy the linear wave equation:

$$\Delta \psi = \frac{1}{v^2} \frac{\partial^2 \psi}{\partial t^2} . \tag{2.1}$$

Here, we are using the Euclidian metric on the space of $x^i$ (which will be $\mathbb{R}^3$), namely:

$$g = \delta_{ij} \, dx^i \, dx^j , \tag{2.2}$$

in order to define the Laplacian operator, which is then:

$$\Delta = \delta^{ij} \partial_{ij} = \delta_{ij} \frac{\partial^2}{\partial x^i \partial x^j}, \tag{2.3}$$

in which $\delta^{ij}$ is the inverse matrix to $\delta_{ij}$, although both matrices will have the same components.

One can also think of $g$ as something that defines a scalar product on the vector space $\mathbb{R}^3$ {$(x^1, x^2, x^3)$} and its dual vector space $\mathbb{R}^{3*}$, which consists of linear functionals on the vectors in $\mathbb{R}^3$, or *1-forms*. Hence, if $\mathbf{v} = (v^1, v^2, v^3)$ and $\mathbf{w} = (w^1, w^2, w^3)$ are any vectors in $\mathbb{R}^3$ and $\alpha = (\alpha_1, \alpha_2, \alpha_3)$ and $\beta = (\beta_1, \beta_2, \beta_3)$ are any 1-forms in $\mathbb{R}^{3*}$ then the scalar products will be defined by:

$$< \mathbf{v}, \mathbf{w} > = \delta_{ij}\, v^i v^j, \qquad < \alpha, \beta > = \delta^{ij}\, \alpha_i \beta_j . \tag{2.4}$$

*a. Time-harmonic solutions.* – A wave function $\psi$ is said to be *time-harmonic* [**3-5**] if it can be expressed in the form:

$$\psi(t, x^i) = u(x^i)\, e^{i\omega t}, \tag{2.5}$$

in which $u$ is a complex-valued spatial function that describes the "shape" of the wave form, while the exponential factor imposes a sinusoidal variation of that shape with an angular frequency of $\omega$. Since $\omega$ is assumed to be constant, one also refers to such waves as *monochromatic*.

Basically, one is separating the variables into time and space variables, with a separation constant of $\omega$. Substituting the form (2.5) into the wave equation (2.1) and cancelling the exponential factor on both sides of the resulting equation gives the *Helmholtz equation:*

$$\Delta u + \kappa^2 u = 0, \tag{2.6}$$

in which the wave number $\kappa = \kappa(x^i)$ that is associated with $\omega$ is defined by:

$$\kappa = \frac{\omega}{v} = \frac{2\pi}{\lambda}, \tag{2.7}$$

where $\lambda$ is the wavelength that is associated with $\kappa$ ([1]). Notice that although $\omega$ is a constant (viz., a separation constant), by definition, $\kappa$ does not have to be constant since that would depend upon the optical properties of the medium in which the wave propagates.

*b. Introduction of polar coordinates on the complex plane.* – Since the stationary wave function $u$ is still complex-valued, it would be useful to introduce polar coordinates $(a, \theta)$ into the complex number plane and see how the Helmholtz equation splits into real and imaginary equations. Namely, one sets:

$$u(x) = a(x)\, e^{i\theta(x)}, \tag{2.8}$$

---

([1]) Strictly speaking the concept of "wavelength" is well-defined only for spatially-periodic wave functions, such as plane waves. However, one can still regard it, more generally, as a "characteristic length" for the wave in the case where no such spatial periodicity is present. Similarly, one can regard the wave number of a wave, more fundamentally, as something that describes the "density" of the wave surfaces with respect to phase.

in which $a(x)$ and $\theta(x)$ are real-valued functions of position that one can think of as the *amplitude* and *phase* functions that are associated with $u$. The level surfaces $\theta(x) =$ const. are regarded as the *wave surfaces* or *wave fronts*; they are also sometimes referred to as *isophases*.

Substitution of (2.8) in (2.6) and dropping the exponential factor $e^{i\theta}$ gives:

$$0 = (\Delta a - <d\theta, d\theta> a + \kappa^2 a) + i(\Delta\theta + 2 <d\theta, da>).$$

Since $a$ and $\theta$ were assumed to both be real functions, the vanishing of the right-hand side would imply the simultaneous vanishing of the real and imaginary parts, which would give the pair of equations:

$$<d\theta, d\theta> = \kappa^2 + \frac{\Delta a}{a}, \qquad 0 = \Delta\theta + 2<d\theta, da>. \qquad (2.9)$$

One sees that the 1-form $d\theta$ has an immediate geometric significance as the normal 1-form to the wave surfaces. That is, whenever $d\theta$ is evaluated on a vector **v** that is tangent to a wave surface, the result $d\theta(\mathbf{v}) = v^i \partial_i \theta$ will be zero.

One can use the metric (2.2) on the ambient space in which one finds the optical medium to associate the 1-form $d\theta$ with a vector field $\nabla\theta$, which is simply the *gradient* of the phase function:

$$\nabla\theta = \theta^i \partial_i, \quad \text{where} \quad \theta^i \equiv \delta^{ij} \partial_j \theta. \qquad (2.10)$$

The vector field $\nabla\theta$ is then normal to the wave surface, as long as it does not vanish. Points where $\nabla\theta$ or $d\theta$ vanish will be called *singular points*, and will not be considered in what follows, although they do have a non-trivial significance in optics as focal points of the congruences of *light rays* that are the integral curves to the vector field $\nabla\theta$. That is, a curve $x(\sigma) = (x^1(\sigma), x^2(\sigma), x^3(\sigma))$ in $\mathbb{R}^3$ that is parameterized by $\sigma$ will be an integral curve of the vector field $\nabla\theta$ iff its tangent vector field $dx/d\sigma$ is equal to the vector $\nabla\theta$ at each point:

$$\frac{dx}{d\sigma} = \nabla\theta(x(\sigma)), \qquad (2.11)$$

or in component form:

$$\frac{dx^i}{d\sigma} = \delta^{ij} \partial_j \theta. \qquad (2.12)$$

The support of $\nabla\theta$ is contained in the domain of definition of $\theta$, which coincides with the support of the wave function $\psi$, and the isophase surfaces represent the cross-sections of the congruence of light rays. By definition, the congruence is *surface-normal* then.

As for the second equation in (2.9), if one multiplies both sides of it by the amplitude function $a$ then one will have:

$$a^2 \Delta\theta + 2\, a <d\theta, da> = a^2 \operatorname{div}(\nabla\theta) + da^2(\nabla\theta) = \operatorname{div}(a^2 \nabla\theta),$$

so if one defines the vector field:

$$\mathbf{J} = a^2 \nabla\theta \qquad (2.13)$$

then the equation for the amplitude will take the form of the vanishing of the divergence of a certain vector field:

$$0 = \operatorname{div} \mathbf{J}. \qquad (2.14)$$

That tends to suggest that the vector field $\mathbf{J}$ describes some sort of conserved current that is associated with the motion. Note that since $\theta$ is dimensionless, its gradient will have the dimension of inverse length, and the dimension of $\mathbf{J}$ will be the square of the dimension of the amplitude over length. However, from now, on our primary focus shall be the first of equations (2.9), since that is the one leads to the geodesic equations.

**3. The geometrical optics approximation.** – Traditionally, when one has geometrical optics in mind, one first introduces the *index of refraction* of the medium [1]:

$$n = \frac{c}{v}, \qquad (3.1)$$

in which $c$ is a constant that describes the speed of propagation of electromagnetic waves in the classical electromagnetic vacuum. Hence, $c$ represents a sort of asymptotic ideal optical medium for wave propagation, namely, a linear, isotropic, homogeneous medium that is free from dispersion in the sense of non-local excitations of the medium. That medium (viz., the classical electromagnetic vacuum) is characterized by two constants, namely, the dielectric constant $\varepsilon_0$ and the magnetic permeability $\mu_0$, which are coupled to $c$ by the relationship:

$$c^2 = \frac{1}{\varepsilon_0 \mu_0}. \qquad (3.2)$$

One can then express the wave number $\kappa$ in the form:

$$\kappa = \kappa_0\, n, \qquad \kappa_0 \equiv \frac{\omega}{c}. \qquad (3.3)$$

Hence, $\kappa_0$ is a constant that one can interpret as the wave number that is associated with $\omega$ for waves that propagate in the classical vacuum.

---

[1] The fact that we are defining an index of refraction that does not depend upon the direction of propagation implies that we are restricting ourselves to an *isotropic* medium.

Next, one factors the phase function $\theta$ into the product:

$$\theta = \kappa_0 S, \qquad (3.4)$$

in which the function $S(x)$ that has been introduced is called the *eikonal*. Since the phase function $\theta$ is dimensionless (i.e., units of radians) and $\kappa_0$ has a dimension of 1 / length, the value of $S$ will have a dimension of length. Indeed, when one starts from an action functional for light rays (viz., *Fermat's principle*), the function $S$ will take the form of a two-point function $S(x_1, x_2)$ of position that comes about when one assumes the that arc-length functional that associates any path segment between the points $x_1$ and $x_2$ with the length of that path segment assigns the same length to all paths. Since that assignment sounds a bit coarse-grained as a measure of path length, one must realize that it only becomes useful in the geometrical optics *approximation*.

So far, we have not introduced any actual approximation, but only a redefinition of $\theta$. In order to see where the approximation comes in, we substitute (3.3) and (3.4) in the first equation in (2.9) and divide by $\kappa_0$:

$$<dS, dS> = n^2 + \frac{1}{\kappa_0^2} \frac{\Delta a}{a}. \qquad (3.5)$$

When we consider the *high-frequency* (i.e., *short-wavelength*) approximation that lets $\kappa_0$ become infinite, what we will be left with is:

$$<dS, dS> = n^2. \qquad (3.6)$$

This is the celebrated *eikonal* equation, which goes back to Hamilton [7] and Jacobi [8], in principle, although the terminology is due to Bruns [9]. One sees that in that high-frequency limit, which is the *geometrical optics* approximation, the equations for the real and imaginary parts of the Helmholtz equation will be decoupled, so one can, in principle, solve the eikonal equation for the function $S$ (which will then give the phase function $\theta$) without having to know what the corresponding amplitude function $a$ is. However, as will be shown, that is also the point at which one is ruling out diffraction for the propagating wave.

One might see a close analogy between the relationship of the eikonal $S$ to the amplitude $a$ and the relationship of the period (or frequency) of a pendulum to its amplitude. That is, for the simple pendulum (which is a small-amplitude approximation), the period is independent of the amplitude, but as the approximation breaks down at larger amplitudes the period becomes increasingly dependent upon the amplitude.

Another way of characterizing the geometrical optics approximation is that one is only dealing with the behavior of individual light rays, or at best an "infinitely-thin" congruence of light rays that is centered on one of them. Hence, if one drops the approximation then one will need to consider congruences of finite cross-section, which suggests that one is not associating the propagating wave front with a point-particle that moves along a single integral curve for its normal vector field, but a congruence of curves that describes the motion of something that occupies a finite volume of space; i.e., extended matter.

**4. Derivation of the geodesic equations for diffracted light rays.** – Now let us return to the basic set of equations (2.9) for the amplitude and phase. In order to show how those equations lead to the geodesic equations for spatial light rays, we shall first convert the first of those equations into a different form that has slightly more physical significance.

*a. Conversion of amplitude and phase equations.* – We shall continue to use the definitions of $dS$ and $\kappa_0$ as in (3.4), so the phase equation will take the form (3.5). However, we shall define a new index of refraction that includes the perturbation of the one that is defined by the optical medium (viz., $n$) by the amplitude $a$ of the wave that passes through it:

$$\bar{n}^2(x,a) = n^2 + \frac{1}{\kappa_0^2}\frac{\Delta a}{a}. \tag{3.7}$$

Hence, we can now write the perturbed eikonal equation in the form:

$$<dS, dS> = \bar{n}^2. \tag{3.8}$$

One can now follow the usual route that one employs for the unperturbed index $n$ and say that the latter equation merely says that the length of the 1-form $dS$ or the vector field $\nabla S$ is given by the perturbed index. Thus, if $s$ is the arc-length parameter of a curve $x(s)$, as defined by the ambient metric (so the speed of the curve will be unity), and the curve is an integral curve of $\nabla S$ then one must have:

$$\frac{dx}{ds} = \frac{1}{\bar{n}}\nabla S(x(s)) \quad \text{or} \quad \nabla S = \bar{n}\frac{dx}{ds}. \tag{3.9}$$

In components that is:

$$\frac{dx^i}{ds} = \frac{1}{\bar{n}}\delta^{ij}\partial_j S \quad \text{or} \quad \partial_i S = \bar{n}\,\delta_{ij}\frac{dx^j}{ds}. \tag{3.10}$$

Now:

$$\frac{d}{ds}(dS) = i_t\, d^2 S = \frac{1}{\bar{n}}<dS, d^2S>, \tag{3.11}$$

whose component form is:

$$\frac{d}{ds}(\partial_i S) = \frac{dx^i}{ds}\partial_i S = \frac{1}{\bar{n}}\delta^{jk}\partial_j S\,\partial_{ki}S. \tag{3.12}$$

On the other hand, differentiating (3.8) gives:

$$d<dS, dS> = 2<dS, d^2S> = 2\bar{n}\,d\bar{n}, \quad \text{so} \quad \frac{1}{\bar{n}}<dS, d^2S> = d\bar{n},$$

and when this is combined with (3.9) and (3.11), that will give the basic equation of the light rays:

$$\frac{d}{ds}\left(\bar{n}\frac{dx}{ds}\right) = \nabla \bar{n} \ . \tag{3.13}$$

In components, that takes the form:

$$\frac{d}{ds}\left(\bar{n}\frac{dx^i}{ds}\right) = \delta^{ij}\partial_j\bar{n} \ . \tag{3.14}$$

In the geometrical optics approximation, the perturbed index will converge to the usual one *n*, and the latter equation will become the usual one that one encounters in gradient-index optics [**10**].

As for the second equation in (2.9), if one replaces $d\theta$ with $\kappa_0 \, dS$, cancels the common factor of $\kappa_0$, and multiplies both sides by the amplitude function *a* then one will have:

$$a^2 \, \Delta S + 2 \, a < dS, da > = a^2 \, \mathrm{div}\,(\nabla S) + da^2 \, (\nabla S) = \mathrm{div}\,(a^2\,\nabla S) \ ,$$

so if one defines the vector field:

$$\mathbf{J}_S = a^2\,\nabla S = \frac{1}{\kappa_0}\mathbf{J} \tag{3.15}$$

then the equation for the amplitude will take the form of the vanishing of the divergence of a certain vector field:

$$0 = \mathrm{div}\,\mathbf{J}_S \ . \tag{3.16}$$

Therefore, the vector field $\mathbf{J}_S$ will describe a conserved current that is associated with the motion, and its dimension will be the square of the dimension of the amplitude.

*b. Deriving the geodesic equations.* – Now, assume that the ambient metric is Euclidian, so we can focus on the contribution to the geodesic equations that come from the index of refraction. First, let us expand the left-hand side of (3.13):

$$\frac{d\bar{n}}{ds}\frac{dx}{ds} + \bar{n}\frac{d^2x}{ds^2} = d\bar{n}$$

or

$$0 = \frac{d^2x}{ds^2} + \frac{1}{\bar{n}}\left(\frac{d\bar{n}}{ds}\frac{dx}{ds} - d\bar{n}\right) \ .$$

Now, since:

$$\frac{d\bar{n}}{ds} = \partial_i\bar{n}\frac{dx^i}{ds} \qquad \text{and} \qquad \delta_{jk}\frac{dx^j}{ds}\frac{dx^k}{ds} = 1 \ ,$$

that will make:

$$0 = \frac{d^2x}{ds^2} + \frac{1}{\bar{n}}(\partial_j \bar{n}\, \delta^i_k - \partial^i \bar{n}\, \delta_{jk})\frac{dx^j}{ds}\frac{dx^k}{ds} \ .$$

The second term on the right-hand side needs to be symmetrized in $j$ and $k$, since it is being contracted with a product of velocities that is symmetric in those indices:

$$0 = \frac{d^2x^i}{ds^2} + \frac{1}{2\bar{n}}(\partial_j \bar{n}\, \delta^i_k + \partial_k \bar{n}\, \delta^i_j - 2\partial^i \bar{n}\, \delta_{jk})\frac{dx^j}{ds}\frac{dx^k}{ds}. \tag{3.17}$$

This equation already looks vaguely reminiscent of the geodesic equation, and all that is necessary to put it into a more familiar form is to first define the conformally-transformed metric:

$$\bar{g} = \bar{n}^2 g, \tag{3.18}$$

where we are temporarily reverting to a more general ambient metric $g$, and then compute the components of the Levi-Civita connection that goes with the transformed metric:

$$\bar{\Gamma}^i_{jk} = \tfrac{1}{2}\bar{g}^{il}(\partial_j \bar{g}_{lk} + \partial_k \bar{g}_{lj} - \partial_l \bar{g}_{jk}). \tag{3.19}$$

Since:

$$\bar{g}^{il} = \frac{1}{\bar{n}^2} g^{il} \qquad \text{and} \qquad \partial_j \bar{g}_{lk} = 2\bar{n}\, \partial_j \bar{n}\, g_{lk} + \bar{n}^2 \partial_j g_{lk},$$

one finds that:

$$\bar{\Gamma}^i_{jk} = \Gamma^i_{jk} + \bar{\tau}^i_{jk}, \tag{3.20}$$

in which:

$$\Gamma^i_{jk} = \tfrac{1}{2} g^{il}(\partial_j g_{lk} + \partial_k g_{lj} - \partial_l g_{jk}) \tag{3.21}$$

are the components of the Levi-Civita connection that goes with $g$, and:

$$\bar{\tau}^i_{jk} = \bar{\Gamma}^i_{jk} - \Gamma^i_{jk} = \tfrac{1}{2}(\partial_j \bar{n}\, \delta^i_k + \partial_k \bar{n}\, \delta^i_j - \partial^i \bar{n}\, g_{jk}) \tag{3.22}$$

are the components of the difference 1-form that takes the Levi-Civita connection of $g$ to the Levi-Civita connection of the conformally-transformed metric $\bar{g}$, namely:

$$\bar{\tau}^i_j = \bar{\tau}^i_{jk}\, dx^k. \tag{3.23}$$

In particular, when the ambient space in which the waves propagate is Euclidian space, so $g_{ij} = \delta_{ij}$, one will have simply:

$$\bar{\Gamma}^i_{jk} = \bar{\tau}^i_{jk}. \tag{3.24}$$

When one forms the geodesic equations relative to the conformally-transformed metric, one must keep in mind that geodesics must minimize the arc-length that is defined by the metric, so when one changes the metric, one must also change the arc-length parameter accordingly. Hence, one can say that the new arc-length parameter $\bar{s}$ satisfies:

$$d\bar{s} = \bar{n}\, ds \quad \text{or} \quad \frac{d\bar{s}}{ds} = \bar{n}, \quad \frac{d}{ds} = \bar{n}\frac{d}{d\bar{s}}. \tag{3.25}$$

That will then make:

$$\frac{dx^i}{ds} = \bar{n}\frac{dx^i}{d\bar{s}} \quad \text{and} \quad \frac{d^2x^i}{ds^2} = \bar{n}^2\left(\frac{d^2x^i}{d\bar{s}^2} + \frac{1}{\bar{n}}\partial_{(j}\bar{n}\,\delta^i_{k)}\frac{dx^j}{d\bar{s}}\frac{dx^k}{d\bar{s}}\right), \tag{3.26}$$

in which the parentheses in the subscripts indicate that one must symmetrize those two indices.

When one makes those substitutions in (3.17), one will get:

$$0 = \bar{n}^2\left[\frac{d^2x^i}{d\bar{s}^2} + \bar{\tau}^i_{jk}\frac{dx^j}{d\bar{s}}\frac{dx^k}{d\bar{s}}\right],$$

so since one is assuming the special case of a conformally-deformed Euclidian metric, (3.24) will apply, and one will conclude that:

$$0 = \frac{d^2x^i}{d\bar{s}^2} + \bar{\Gamma}^i_{jk}\frac{dx^j}{d\bar{s}}\frac{dx^k}{d\bar{s}}, \tag{3.27}$$

which is, in fact, the form of the geodesic equations for the conformally-transformed metric.

*c. Decomposing the connection into ambient and diffracted parts.* – Recall the expression (3.7) for the perturbed index of refraction and now express it in the form:

$$\bar{n}^2 = n^2 + n_a^2, \quad \text{with} \quad n_a^2 \equiv \frac{1}{\kappa_0^2}\frac{\Delta a}{a}. \tag{3.28}$$

Differentiating this will give:

$$d\bar{n} = \frac{n}{\bar{n}}dn + \frac{n_a}{\bar{n}}dn_a, \tag{3.29}$$

which essentially amounts to a convex combination of the differentials, in the sense that:

$$\left(\frac{n}{\bar{n}}\right)^2 + \left(\frac{n_a}{\bar{n}}\right)^2 = \frac{n^2 + n_a^2}{\bar{n}^2} = \frac{n^2 + n_a^2}{n^2 + n_a^2} = 1.$$

When one takes (3.29) into account while computing the components of the difference 1-form, one will find that:

$$\bar{n}\,\bar{\tau}^i_{jk} = \partial_j \bar{n}\,\delta^i_k - \partial_k \bar{n}\,\delta^i_j = \frac{n}{\bar{n}}(\partial_j n\,\delta^i_k + \partial_k n\,\delta^i_j - \partial^i n\,\delta_{jk}) + \frac{n_a}{\bar{n}}(\partial_j n_a\,\delta^i_k + \partial_k n_a\,\delta^i_j - \partial^i n_a\,\delta_{jk})\,,$$

which one can write in the form:

$$\bar{\tau}^i_{jk} = \frac{n^2}{\bar{n}^2}\tau^i_{jk} + \frac{n_a^2}{\bar{n}^2}\hat{\tau}^i_{jk}, \tag{3.30}$$

with:

$$\tau^i_{jk} \equiv \frac{1}{n}(\partial_j n\,\delta^i_k + \partial_k n\,\delta^i_j - \partial^i n\,\delta_{jk}), \tag{3.31}$$

$$\hat{\tau}^i_{jk} \equiv \frac{1}{n_a}(\partial_j n_a\,\delta^i_k + \partial_k n_a\,\delta^i_j - \partial^i n_a\,\delta_{jk}). \tag{3.32}$$

Hence, the Levi-Civita connection of the ambient space has been perturbed by both the "classical" index of refraction $n$ and the "diffracted" one $n_a$ by way of additive contributions:

$$\bar{\Gamma}^i_{jk} = \Gamma^i_{jk} + \frac{n^2}{\bar{n}^2}\tau^i_{jk} + \frac{n_a^2}{\bar{n}^2}\hat{\tau}^i_{jk}. \tag{3.33}$$

Therefore, the equations of the geodesics of the metric that is perturbed in both ways can be written in the form:

$$\frac{d^2 x^i}{d\bar{s}^2} + \left(\Gamma^i_{jk} + \frac{n^2}{\bar{n}^2}\tau^i_{jk}\right)\frac{dx^j}{d\bar{s}}\frac{dx^k}{d\bar{s}} = -\frac{n_a^2}{\bar{n}^2}\hat{\tau}^i_{jk}\frac{dx^j}{d\bar{s}}\frac{dx^k}{d\bar{s}}. \tag{3.34}$$

In this form, one sees that the only way that $n_a$ contributes to the left-hand side is by way of the perturbed index of refraction, while the right-hand side will vanish with $n_a$, which amounts to vanishing in the geometrical optics approximation (viz., $\kappa_0$ becomes infinite). Thus, one would expect that the non-vanishing of that right-hand side would be characteristic of diffracted geodesics.

One can use the Frenet-Serret equations for the light rays to show that equations (3.34) imply a corresponding equation for the curvature of the light rays, which we temporarily denote by $\kappa(\bar{s})$, although we have been using the symbol $\kappa$ to represent a wave number up to now. Recall [11] that the unit tangent vector field $\mathbf{t}(\bar{s})$ and unit normal vector field $\mathbf{n}(\bar{s})$ relate to the first and second derivatives of position $x(\bar{s})$ with respect to arc-length by:

$$\mathbf{t}(\bar{s}) = \frac{dx}{d\bar{s}}, \quad \kappa(\bar{s})\mathbf{n}(\bar{s}) = \frac{d\mathbf{t}}{d\bar{s}} = \frac{d^2 x}{d\bar{s}^2}. \tag{3.35}$$

When one makes those replacements in (3.34) and takes the scalar product of both sides with $\mathbf{n}(\bar{s})$, one will get:

$$\kappa(\bar{s}) = -n_i \left( \Gamma^i_{jk} + \frac{n^2}{\bar{n}^2} \tau^i_{jk} + \frac{n_a^2}{\bar{n}^2} \hat{\tau}^i_{jk} \right) t^j t^k. \tag{3.36}$$

Thus, the light rays are curved by three additive contributions: The metric of the ambient space, the index of refraction of the medium, and the effects of diffraction.

Since many of the worked problems in diffraction pertain to homogeneous media in Euclidian ambient spaces, it is useful to specialize (3.34) accordingly. Now, the $\Gamma$'s and the $\tau$'s will all vanish for such a medium in such a space, so what will remain of the geodesic equations is:

$$\frac{d^2 x^i}{ds^2} = -\frac{n_a^2}{\bar{n}^2} \hat{\tau}^i_{jk} \frac{dx^j}{ds} \frac{dx^k}{ds}. \tag{3.37}$$

That makes it clear that light rays can still depart from rectilinearity depending upon how the amplitude of the wave propagates.

For the purpose of calculation, one can use (3.32) to reduce (3.37) to the simpler form:

$$\frac{d^2 x^i}{ds^2} = -\frac{1}{\bar{n}^2} \left( \frac{dn_a^2}{ds} \frac{dx^i}{ds} - \partial^i n_a^2 \right). \tag{3.38}$$

Since this can be expressed in the form:

$$\bar{n}^2 \frac{d^2 x^i}{ds^2} + \frac{dn_a^2}{ds} \frac{dx^i}{ds} = -\partial^i n_a^2,$$

the fact that $n$ is constant will imply that:

$$\frac{d\bar{n}^2}{ds} = \frac{dn_a^2}{ds},$$

so one ultimately has:

$$\frac{d}{ds}\left( \bar{n}^2 \frac{dx^i}{ds} \right) = \partial^i n_a^2. \tag{3.39}$$

That is the basic equation of gradient-index optics (3.14) for a homogeneous optical medium that is perturbed by diffraction, so the index gradient in question is due to only the existence of diffraction. One cautiously proposes that to some extent the effect of *diffraction* on light *waves* is equivalent to an effective *refraction* of light *rays* that is due to the appearance of $dn_a \neq 0$ in regions where diffraction takes place, such as apertures. An essential difference though is that refraction at an interface between homogeneous optical media will typically bend a congruence of parallel light rays into another congruence of parallel light rays, while typically diffraction bends a congruence of parallel light rays into a cone of light rays that are no longer parallel. One might explain that by observing that $n_a$ will typically vary across an aperture in such a way that the angle through which the light ray is bent will vary from zero at the center to something larger at the boundary of the aperture, and the rate of change of that angle will also be greatest near that boundary. More to the point, the angle (as well as $dn_a$) will be non-zero only within a distance from the boundary that is less than say, one wavelength.

**5. Obtaining the explicit form for the diffraction corrections**. – In order to find the solutions to the equations of diffracted geodesics, it helps to know the explicit form of the amplitude function $a(x)$ beforehand. One can then calculate an explicit form for $n_a = n_a(x)$ from that and see what form the geodesic equations will take in each case.

In order to obtain $a(x)$, one must return to the traditional methods of the theory of diffraction (cf., e.g., [**3-5**]), which are indirectly related to finding solutions to various boundary-value problems in the Helmholtz equation. That is, one considers the problem of finding $u(x)$ for a particular aperture, such as a circular hole in an opaque planar screen or its complementary arrangement of an opaque planar disc in a transparent medium or a slit in a planar screen. Once one knows $u(x)$ for the configuration, one can obtain $a(x)$ from its amplitude, i.e.:

$$a = \| u \| = \sqrt{u u^*} . \tag{4.1}$$

Hence, we shall briefly recall the nature of the boundary-value problems that are defined by optical scenarios in which diffraction is present. One starts with Green's formula, which says that for any twice-continuously differentiable functions $f$ and $g$ on a region of space $V$ with a boundary $\partial V$, one must have:

$$\int_V (\Delta f) g \, dV - \int_V f \Delta g \, dV = \int_{\partial V} \left( g \frac{\partial f}{\partial n} - f \frac{\partial g}{\partial n} \right) dS , \tag{4.2}$$

which becomes:

$$\int_V (\Delta_\kappa f) g \, dV - \int_V f \Delta_\kappa g \, dV = \int_{\partial V} \left( g \frac{\partial f}{\partial n} - f \frac{\partial g}{\partial n} \right) dS , \tag{4.3}$$

in the case of the Helmholtz operator:

$$\Delta_\kappa = \Delta + \kappa^2 \tag{4.4}$$

since the algebraic parts of $(\Delta_\kappa f) g$ and $f (\Delta_\kappa g)$ cancel.

One symbolically introduces the Dirac delta function $\delta(x, x')$ (which is not really a function, but is defined only in terms of distributions) and defines the following boundary value problem for the Green function $G(x, x')$:

$$\Delta'_\kappa G(x, x') = -\delta(x, x'), \qquad G(x, x') = 0, \text{ for } x' \in \partial V. \tag{4.5}$$

Sommerfeld [3] also imposes a "radiation condition" on the Green function that is based upon its presumed behavior at infinity, namely, that if $r$ is the radius of a large (but not infinite) sphere about a point-like light source then one must have:

$$\lim_{r \to \infty} \left[ r \left( \frac{\partial G}{\partial n} - i\kappa G \right) \right] = 0. \tag{4.6}$$

If we specialize (4.3) to the boundary-value problem for the Helmholtz equation by setting $f = u$, which is a solution of the Helmholtz equation, and $g = G$ then we get:

$$u(x) = \int_{V'} \delta(x, x') u(x') \, dV' = -\int_{\partial V} u(x') \frac{\partial G}{\partial n'} \, dS'. \tag{4.7}$$

Hence, in theory, as long as one has found a Green function for the given boundary-value problem in the Helmholtz problem, one can calculate the solution $u(x)$ from its assumed values on that boundary.

The explicit form of the Green function amounts to Huygens's principle [12], namely, that each point $x'$ in the aperture can be regarded as a source of a spherical wave whose wave function as a function of $x$ can be represented in the form:

$$G(x, x') = G(x - x') = \frac{e^{i\kappa \|x - x'\|}}{\|x - x'\|}, \tag{4.8}$$

in which:

$$\|x - x'\| = \sqrt{(x^1 - x'^1)^2 + (x^2 - x'^2)^2 + (x^3 - x'^3)^2} \tag{4.9}$$

is the Euclidian distance between the two points.

One can then write (4.7) in the form:

$$u(x) = -\int_{\partial V} u(x') \frac{\partial}{\partial n'} \left( \frac{e^{i\kappa \|x - x'\|}}{\|x - x'\|} \right) dS'. \tag{4.10}$$

In principle, it would seem that the problem of finding $a(x)$ has been solved by this. However, when one gets into the literature of diffraction [3-5], one eventually finds that the complexity involved with solving a boundary-value problem in the Helmholtz equation makes that problem something that is usually only amenable to computer modelling. There are also techniques that involve asymptotic expansions of the wave function [5]. Typically, what one obtains (see Sommerfeld [3]) are approximate formulas for the diffracted wave function along – say – the center line of the aperture, but without actually obtaining a closed-form solution for the wave function $u(x)$ in the rest of space, and thus for $a(x)$, which would be the starting point for calculating the diffraction corrections to the geodesic equations for a specific optical configuration.

Hence, one can at best, give the function $a(x)$ properties that are reasonable under the circumstances and see how that affects the geodesics that represent light rays. For instance, one might assume that $a(x)$ is constant over most of the aperture and changes only close to its boundary, e.g., within a wavelength of it. Of course, that analysis would be quite involved in its own right, so it shall be postponed for a later study.

**6. Concluding remarks.** – The basic results of this study have been to extend the usual geodesic equations for light rays, which typically assume the geometrical optics approximation, to equations that would include a contribution from diffraction effects and to show how the Levi-Civita connection that enters into those equations decomposes into a sum of contributions from the ambient space metric, the index of refraction of the optical medium, and the diffraction effects.

The previous section of the article made it clear that in order to proceed with any analysis of diffracted geodesics for the usual optical situations in which diffraction is treated, one would have to have an explicit form for the amplitude of the propagating wave as a function of position in space. However, that is a much deeper problem than it might first appear to be, although the most promising path to go down would probably be that of wave functions that are represented by asymptotic expansions.

Another direction to pursue is the closely-related world of quantum wave mechanics. Indeed, the stationary Schrödinger equation, as well as the stationary Klein-Gordon equation, also takes the form of Helmholtz's equation, so the applicability of the mathematical methods would be direct. Presumably, it might lead to a different perspective on wave-particle duality and the optical-mechanical analogy, as well as a different perspective on electron diffraction, in particular.

Also in the context of quantum mechanics, one finds that just as the stationary Schrödinger equation has the Helmholtz form, similarly, the introduction of polar coordinates into the complex plane in which the wave function $u(x)$ takes its values is also the basis for the Madelung-Takabayasi interpretation of wave mechanics [**13**], which is often referred to as the "hydrodynamical" interpretation. The analogue of what we have been calling $n_a^2$ for the optical interpretation is referred to as the "quantum potential" in that case, although the role of $1/\kappa_0$ is played by $\hbar$ in quantum mechanics. The asymptotic expansions that model the diffraction effects in optics become the WKB approximation in quantum mechanics. In effect, one then has two analogies to pursue in quantum theory: the optical-mechanical analogy and the continuum-mechanical analogy.

_________